\documentclass[prd]{revtex4}
\usepackage{epsfig}
\usepackage{t1enc}
\usepackage[latin1]{inputenc}

\usepackage{epsf,amsfonts,amssymb,epsfig,bm,amsmath,yfonts}
\usepackage{graphicx}
\newcommand{\eqn}[1]{(\ref{#1})}
\newcommand{\be}{\begin{equation}}
\newcommand{\ee}{\end{equation}}
\newcommand{\ben}{\begin{displaymath}}
\newcommand{\een}{\end{displaymath}}
\newcommand{\bea}{\begin{eqnarray}}
\newcommand{\eea}{\end{eqnarray}}
\newcommand{\bean}{\begin{eqnarray*}}
\newcommand{\eean}{\end{eqnarray*}}

\newcommand{\ba}{\begin{array}}
\newcommand{\ea}{\end{array}}
\newcommand{\bi}{\begin{itemize}}
\newcommand{\ei}{\end{itemize}}

\def\G {\Gamma}

\def\e {\epsilon}

\def\otaula{\begin{tabular}}
\def\ctaula{\end{tabular}}

\renewcommand{\t}{\theta}

\def\CR{\mathbb{R}}
\def\CM{\mathcal{M}}

\def\8M{$\CM_8$}

\def\be{\begin{equation}}
\def\ee{\end{equation}}
\def\G{\Gamma}

\def\ei{e^{\underline{i}}}

\def\e1{e^{\underline{1}}}
\def\1u{\underline{1}}
\def\2u{\underline{2}}

\def\0u{\underline{0}}
\def\e{\epsilon}
\def\target{$\CR^{1,1}\times \mathcal{M}_8$ }
\def\target2{$\CR^{1,1}\times \mathcal{M}_8$,}
\def\9G{\G_{\underline{9}}}

\newcommand{\caln}{\mbox{${\cal N}$}}

\def\nc {N_\mt{c}}
\def\nf {N_\mt{f}}

\def\t6 {T_\mt{D6}}

\newcommand{\mt}[1]{\textrm{\tiny #1}}

\newcommand{\uem}{U(1)_\mt{EM}}
\newcommand{\jem}{J^\mt{EM}}
\newcommand{\A}{{\cal A}}

\begin{document}

\title{Hadron production in electron-positron annihilation computed from the gauge gravity correspondence}

\author{Leonardo Patiño}
\affiliation{Departamento de Física, Facultad de Ciencias, Universidad Nacional Autónoma de México, México D.F. 04510, México}
\author{G. Toledo Sánchez}
\affiliation{Instituto de Física, Universidad Nacional Autónoma de México, México D.F. 04510, México}

\date{\today}

\begin{abstract}
We provide a non-perturbative expression for the hadron production in electron-positron annihilation at zero temperature in a strongly coupled, large-$\nc$ $SU(\nc)$ field theory with $\nf \ll \nc$ quark flavors. The resulting expressions are valid to leading order in the electromagnetic coupling constant but non-perturbatively in the $SU(\nc)$ interactions and the mass of the quark. We obtain this quantity by computing the imaginary part of the hadronic vacuum polarization function $\Pi_q$ using holographic techniques, providing an alternative to the known method that uses the spectrum of infinitely stable mesons determined by the normalizable modes of the appropriated fields in the bulk. Our result exhibits a structure of poles localized at specific real values of $q^2$, which coincide with the ones found using the normalizable modes, and extends it offering the unique analytic continuation of this distribution to a function defined for values of $q^2$ over the complex plane. This analytic continuation permits to include a finite decay width for the mesons. By comparison with experimental data we find qualitatively good agreement on the shape of the first pole, when using the $\rho$ meson parameters and choosing a proper normalization factor. We then estimate the contribution to the anomalous magnetic moment of the muon finding an agreement within 25\%, for this choice of parameters.
\end{abstract}

%\keywords{D-branes, Supersymmetry and Duality, Gauge-gravity correspondence}

%\pacs{11.25.Tq,11.15.Tk,13.66.Bc}

%\preprint{}

\maketitle

\section{Introduction}
Observables like the electromagnetic coupling constant, $\alpha(s)$, and the anomalous magnetic moment of the muon have been accurately measured \cite{Mohr:2008fa,BNLE821}, and have proven to be very sensitive to the effects of Quantum Chromodynamics (QCD), which enter through its contribution to the vacuum polarization \cite{eedata,g-2review}. A standard quantity used to parametrize these QCD effects is the ratio between the following cross sections,
\be
R(s)\equiv \frac{\sigma(e^+ e^- \rightarrow \ hadrons)}{\sigma(e^+ e^- \rightarrow \mu^+ \mu^-)}
\ee
which is related to the vacuum polarization $\Pi(s)$ by
\be
R(s)=12\pi{\mathrm{Im}}\Pi_q(s+i\epsilon). \label{Rpi}
\ee
At large momenta perturbative QCD is justified, and the corrections to $R(s)$ up to  second order in the strong coupling constant $\alpha_s$ have been computed \cite{chetyrkin}. On the other hand, in the low energy regime, QCD becomes a non-perturbative theory and no {\em ab initio} QCD calculations for $R(s)$ exist. Therefore, the experimental information on $R(s)$  in this regime is plugged to built  the theoretical predictions for  $\alpha(s)$ and the anomalous magnetic moment of the muon. This is crucial in the case of the muon anomalous magnetic moment, where the energy dependence of the electromagnetic kernel  enhances the low energy QCD contribution. In fact, the theoretical prediction uncertainty is dominated by this so-called hadronic contribution and, at the moment, a discrepancy with the experimental measurement persist \cite{davier0906.5443}.
Therefore, at low energies, a non-perturbative calculation on $\alpha_s$, accounting for the full dependence of $R(s)$ on the quark mass, is mandatory. 

It is in this context that the gauge/gravity correspondence turns out to be a good alternative as a computational method.
The dual theory of QCD is still unknown, but we can do calculations in $SU(\nc)$ super Yang-Mills (SYM) theories coupled to $\nf$ flavors of fundamental matter whose bare mass is an adjustable parameter. These theories are very different from QCD, but the contribution to $R(s)$ we want to investigate comes from the strongly coupled part of QCD. Although the degrees of freedom differ from those of the theory we are working with, this is also a strongly coupled theory, and one may hope that certain properties of QCD in this regime could be observed in this setting. This supposition has already been used in other contexts to model hadronization \cite{Evans:2007sf}, to find the thermal hadron spectrum \cite{Hatta:2008qx} and to study extensively the strongly coupled quark-gluon plasma \cite{hydro2}.

In this work we compute the hadronic vacuum polarization in this framework at zero temperature, compare its behavior with experimental data, and give an estimate of its contribution to the anomalous magnetic moment of the muon. Our results are valid to leading order in the electromagnetic coupling constant, but are non-perturbative in $\alpha_s$ and the mass of the quark. 

\section{$R(s)$ in the gauge theory}
We will obtain $R(s)$ from its expression in terms of the hadronic part of the vacuum polarization function  $\Pi_q(q^2)$, which at zero temperature is related to its tensorial form $\Pi^{\mu\nu}_q(q)$ by
\be
\Pi_q(q^2)=\frac{{\mathrm{Tr}}(\Pi^{\mu\nu}_q(q))}{{q^2d}},\label{Pi}
\ee
where $\mu$ and $\nu$ run from $0$ to $d$, and $d+1$ is the number of dimensions of the gauge theory we are working in.

The matter content of the theories we are working with consists of fundamental and adjoint fields. The former include $\nf$ flavors of fundamental fermions $\Psi^a$ and scalars $\Phi^a$, $a=1, \ldots, \nf$, to which we will refer indistinctly as `quarks'. To be as close as possible to QCD, we will couple only the fundamental fields to electromagnetism and set the electric charge of the adjoint fields to zero, so they do not affect our results for $R(s)$. For this we introduce an additional, dynamical, Abelian gauge field (the photon) that couples with strength (electric charge) $e$ to the fundamental fields. This amounts to replacing the $SU(\nc)$-covariant derivative $D_\mu$ by  ${\cal D}_\mu = D_\mu - i e \A_\mu$ when acting on the fundamental fields, and adding a kinetic term for the photon. In this way we obtain a $SU(\nc) \times \uem$ gauge theory whose Lagrangian is
\be
{\cal L} = {\cal L}_{SU(\nc)} - \frac{1}{4} {\cal F}_{\mu \nu}^2 + e \A^\mu J^\mt{EM}_\mu \,,
\ee
with ${\cal F}_{\mu\nu} = \partial_\mu \A_\nu - \partial_\nu \A_\mu$, and the electromagnetic current given by 
\be
\jem_\mu = \bar{\Psi} \gamma_\mu \Psi + \frac{i}{2} \Phi^* {\cal D}_\mu \Phi 
- \frac{i}{2} \left( {\cal D}_\mu \Phi \right)^* \Phi \,,
\label{current}
\ee
where a sum over flavor and color indices is implicit.

In this theory we compute $\Pi^{\mu\nu}_q(q)$ using the following expression in terms of the correlator of two electromagnetic currents of quarks,
\be
\Pi^{\mu\nu}_q(q) =  i \int d^{d+1} x \, e^{-i q \cdot x} \, \Theta (x^0) 
\langle [ \jem_\mu (x), \jem_\nu (0) ] \rangle .
\label{green}
\ee

Thus in order to study $R(s)$ we must in principle calculate the two-point function in \eqn{green} from the $SU(\nc) \times \uem$ theory. However, as first noticed in \cite{CKMSY}, the terms in the electromagnetic current \eqn{current} proportional to the photon field (implicit in the covariant derivative) lead to higher-order contributions in $e$ for the correlator in \eqn{green}, and can thus be ignored to leading order in $e$. Furthermore we observe that the two-point function of the remaining terms in the current can be calculated in the $SU(\nc)$ theory, since again the effects of including the dynamical photon are of higher order in $e$. We therefore conclude that, to leading order in the electromagnetic coupling constant $e$,  $R(s)$ in an $SU(\nc) \times \uem$ theory is completely determined by the two-point function of the electromagnetic current (\ref{current}) in the $SU(\nc)$ theory. This is the key for us to be able to perform this calculation using the gravitational dual description, since the dual of the $SU(\nc) \times \uem$ is unknown.

\section{Holographic description}
We calculate now this correlator in a ${\caln}=2, SU(\nc)$ SYM theory coupled to $\nf$ flavours of fundamental matter. To this end, we use the gravitational dual of this theory \cite{flavour}, given by a stack of $\nf$ D7 probe branes in the $AdS_5 \times S^5$ background generated by $\nc$ D3 branes in the limit $\nc >> \nf$. For this system we obtain an analytic result for $R(s)$, while for the theory dual to a D4/D6 configuration we could obtain only numerical results, which we do not discuss here since they are qualitatively identical to those of the D3/D7 system. 

In the decoupling limit, the D3-brane solution in the string frame takes the form 
\be
ds^2 = \frac{r^2}{R^2} \left( -dx_0^2 +  d{\bf x}^2 \right) + 
\frac{R^2}{r^2} dr^2 + R^2 d\Omega_{\it 5}^2 \,, 
\label{metricD3} 
\ee
where ${\bf x} = (x^1,x^2,x^3)$ and $R$ is the $AdS_5$ radius given by $R^4=4\pi g_s \nc \ell_s^4$, with $\ell_s$ and $g_s$ the string length and coupling constant, respectively.

According to the gauge/gravity correspondence, string theory on the
background above is dual to $(3+1)$-dimensional, maximally supersymmetric
Yang-Mills theory at zero temperature. Consider now $\nf \ll \nc$ coincident D7-brane probes that share 
the $x^\mu=(x^0,...,x^3)$ directions with the D3-branes, extend along
the radial direction and wrap an $S^3$ inside the $S^5$ of (\ref{metricD3}). If we place the D7-branes at a fixed distance $L$ from the D3-branes in the plane perpendicular to both of them, the configuration will be supersymmetric and this will ensure stability. Under these conditions the Ramond-Ramond field sourced by the D3-branes does not couple to the D7-branes.

In the gauge theory, the D7-branes correspond to introducing $\nf$ 
flavors of fundamental matter \cite{flavour}. These fundamental fields propagate 
along a $(3+1)$-dimensional defect and have the same mass $m_q$ 
dictated by the distance $L$ according to
\be
m_q=\frac{L}{2\pi\ell_s^2}=\frac{L}{R^2}\sqrt{\frac{g_s\nc}{\pi}}.\label{massq}
\ee

The $U(\nf)$ gauge symmetry on the D7-branes is a global flavor symmetry of the gauge theory. If this theory is coupled to electromagnetism as explained above, then the electromagnetic current  $\jem_\mu$ is the conserved current associated to the overall $U(1) \subset U(\nf)$. At strong 't Hooft coupling and large $\nc$, each conserved current of the gauge theory is dual to a gauge field on the gravity side \cite{witten-holo}. To describe the gauge field dual to $\jem_\mu$, let $A_m$, $m=0, \ldots, 7$, be the gauge field associated to the overall $U(1)$ gauge symmetry on the D7-branes. Upon dimensional reduction on the $3$-sphere wrapped by the D7-branes, $A_m$ gives rise to a massless gauge field $\{ A_\mu, A_r \}$, three massless scalars, and a tower of massive Kaluza-Klein 
(KK) modes. All these fields propagate on the five non-compact dimensions of the D7-branes. Following \cite{PSS}, we will work in the gauge $A_r=0$, and furthermore we will set to zero the scalars and the higher KK modes, since they are not of interest here, and unlike in the non-zero temperature case \cite{Mas:2008jz}, this can be done consistently. The gauge field $A_\mu$ is the desired dual to the conserved electromagnetic current $\jem_\mu$ of the gauge theory.

According to the gauge/gravity correspondence, correlation functions of $\jem_\mu$ can be calculated by varying the string partition function with respect to the value of $A_\mu$ at the boundary $r\rightarrow \infty$ of the spacetime  
\eqn{metricD3} \cite{witten-holo}. Under the present circumstances, the string partition function reduces to $e^{iS}$, where $S$ is the sum of the supergravity action and the effective action for the D7-branes. Since $A_\mu$ does not enter the supergravity action, the form of this action will not be needed here. Moreover, the Wess-Zumino part of the D7-branes action does not contribute for the brane orientations and the gauge field polarizations considered in this paper. Therefore, it suffices to consider the Dirac-Born-Infeld part of the D7-branes action, and since we are only interested in the two-point function in \eqn{green}, we keep terms up to quadratic order in the gauge field and use the simplified form of the action
\be
S = - \nf T_\mt{D7} \int_{D7} d^8x \, \sqrt{-\det g} 
\left( 1 + \frac{(2\pi \ell_s^2)^2}{4} F^2 \right) \,.
\label{sdq}
\ee
In this action $F=dA$ is the overall $U(1)$ field strength, $F^2 =g^{kl}g^{mn} F_{km} F_{ln}$, $T_\mt{D7} = 1/(2\pi \ell_s)^7 g_s \ell_s$ is the tension of the D7-branes and $g_{mn}$ is the metric induced over them, given at zero temperature by
\be
ds^2_{D7} = L^2\frac{1+\tilde{r}^2}{R^2} \left( -dx_0^2 +  d{\bf x}^2 \right) + 
\frac{R^2}{1+\tilde{r}^2} d\tilde{r}^2 + \frac{R^2\tilde{r}^2}{1+\tilde{r}^2} d\Omega_{\it 3}^2 \,, 
\label{metricD7} 
\ee
where the radial coordinate over the D7-branes is $\tilde{r}^2=(r^2-L^2)/L^2$.

To get the equations of motion for $A_\mu$ we Fourier-decompose them as
\be
A_\mu(x^0, {\bf x}, r) = \int \frac{d q^0 d^3 {\bf q}}{(2\pi)^4} \, 
e^{-i q^0 x^0 + i {\bf q} \cdot {\bf x}} \, A_\mu (q^0, {\bf q}, r) \,, 
\label{fourier}
\ee
and choose ${\bf q}$ to point in the $x^1$-direction. Under these circumstances the equations of motion for the $A_\mu$-components are

\bea
-\sqrt{-g}g^{11}g^{00}p(w A_1+p A_0)\!+\!\partial_{\tilde{r}}\!\left(\sqrt{-g}g^{\tilde{r}\tilde{r}}g^{00}\partial_{\tilde{r}}A_0\right)&\!\!\!\!=&\!\!\!0 \hspace{.8cm}  \nonumber \\
-\sqrt{-g}g^{11}g^{00}w(w A_1+p A_0)\!+\!\partial_{\tilde{r}}\!\left(\sqrt{-g}g^{\tilde{r}\tilde{r}}g^{11}\partial_{\tilde{r}}A_1\right)&\!\!\!\!=&\!\!\!0 \hspace{.8cm} \label{eomg}\\
-\sqrt{-g}g^{ii}(g^{00}w^2 +g^{11}p^2 )A_i\!+\!\partial_{\tilde{r}}\!\left(\sqrt{-g}g^{\tilde{r}\tilde{r}}g^{ii}\partial_{\tilde{r}}A_i\right)&\!\!\!\!=&\!\!\!0 \hspace{.8cm} \nonumber 
\eea
where we have renamed $q^0$ as $w$, $q^1$ as $p$, and the index $i$ takes values 2 and 3 with no summation implied over it.

For the induce metric \ref{metricD7}, the first and second equations in (\ref{eomg}) imply $A_0=\frac{-p}{w}A_1+\frac{C_1}{\tilde{r}^2}+C_2$, with $C_1$ and $C_2$ integration constants. To apply the method in \cite{SS} and obtain $\Pi_q^{\mu\nu}$ we need the components $A_\mu$ to be regular at $\tilde{r}=0$ and satisfy $\lim_{\tilde{r} \rightarrow \infty}A_\mu =1$. Given that, as we will see below, the non trivial solutions are different from zero at $\tilde{r}=0$, we conclude that $C_1=C_2=0$.

Using $A_0=\frac{-p}{w}A_1$ and (\ref{metricD7}) we see that the equations (\ref{eomg}) can be decoupled from each other and, furthermore, they reduce to the same equation,
\be
\tilde{q}^2\frac{\tilde{r}^3}{(1+\tilde{r}^2)^2}A+\partial_{\tilde{r}}(\tilde{r}^3\partial_{\tilde{r}} A)=0,\label{eomrt}
\ee
where $A$ stands for any of the $A_\mu$'s and we have defined the dimensionless quantity
\be
\tilde{q}^2=\frac{R^4}{L^2}q^2=\frac{g_s\nc}{\pi}\frac{q^2}{{m_q}^2}.\label{qnorm}
\ee

The solution to (\ref{eomrt}) that is, as required above, regular at $\tilde{r}=0$ and satisfies $\lim_{\tilde{r} \rightarrow \infty}A=1$, is uniquely given by
{\small
\be
A=\Gamma\left(\frac{3-\sqrt{1+\tilde{q}^2}}{2}\right)\Gamma\left(\frac{3+\sqrt{1+\tilde{q}^2}}{2}\right)(1+\tilde{r}^2)^{\frac{1-\sqrt{1+\tilde{q}^2}}{2}} F\left(\frac{1-\sqrt{1+\tilde{q}^2}}{2},\frac{3-\sqrt{1+\tilde{q}^2}}{2};2,-\tilde{r}^2\right) \label{solA}
\ee }
where $\Gamma$ and $F$ are, respectively, the Euler gamma function and the standard hypergeometric function.

\subsection{A brief comment on the non-zero temperature case}
In the gauge theory side we know that at zero temperature only $q^2\equiv w^2-p^2$ enters the calculation of the vacuum polarization function, but for the $T\neq 0$ case the solution has $w$ and $p$ separately as relevant parameters. Here we recover this behavior correctly from the gravity side, since from (\ref{eomrt}) it is clear that the only relevant parameter concerning the momentum is $q^2=w^2-p^2$, but for the $T\neq 0$ case, one of the modification to the induce metric $g$ is that $g^{00}\neq -g^{11}$, and it is not difficult to see then that the space of solutions to (\ref{eomg}), and therefore the result for the vacuum polarization function, is parametrized by $w$ and $p$ independently. Since we are working in the $T=0$ case, we don't need to assume anything about the values of $w$ or $p$ independently, nor pick light-like momentum $w=p$ and we can use $q^2$ as our only momentum parameter.

\subsection{Holographic Renormalization}
If we substitute (\ref{metricD7}) and (\ref{solA}) into the action (\ref{sdq}), the integral diverges as we extended it to $\tilde{r} \rightarrow \infty$, which implies that we have to holographically renormalize (\ref{sdq}). It is not difficult to see that the boundary counterterm we need to add is
\be
\left. \delta S =  \nf T_\mt{D7} \!\!\! \int_{\partial D7} \!\!\!\!\!\!\!\!\! d^{4}x\, d\Omega_{\it 3} \, \frac{(2\pi \ell_s^2)^2}{4}R\,\,\,\, {\mathrm{ln}}(\tilde{r}) \sqrt{-\det h}\bar{F}^2 \right|_{\tilde{r}_{max}}
 \,,
\label{dsdq}
\ee
where $\partial$D7 is the hypersurface located at $\tilde{r}_{max}$,  $h_{\alpha\beta}$ the metric induced over it and $\bar{F}^2=h^{\alpha\beta}h^{\gamma\delta}F_{\alpha\gamma}F_{\beta\delta}$. The indices from the beginning of the Greek alphabet run over all the coordinates of the D7-brane except for $\tilde{r}$.

\section{Computation and analysis}
Using (\ref{sdq}) and (\ref{solA}) as an input in the method described in \cite{SS}, we obtain the expression
\be
\Pi_q(\tilde{q}^2) =   -\frac{2 \nf \nc}{3 \tilde{q}^2(2\pi)^8}  \lim_{\tilde{r} \rightarrow \infty}  
\tilde{r}^3 A^*(\tilde{q}^2,\tilde{r}) \partial_{\tilde{r}} A(\tilde{q}^2, \tilde{r}) \,,
\label{corre}
\ee
where we have already taken the trace of $\Pi_q^{\mu\nu}$.

When taking the limit in (\ref{corre}) using the actual expression (\ref{solA}) for $A(\tilde{q}^2, \tilde{r})$, we find a logarithmically divergent term which, as it should be, gets canceled by the contribution of (\ref{dsdq}) to the correlator in (\ref{green}). The reason why we can explicitly check the exact cancellation of the divergences is because in the case we are analyzing, of zero temperature, we have analytic expressions for (\ref{corre}) and (\ref{dsdq}). This procedure yields then a finite result for $\Pi_q(\tilde{q}^2)$, that we substitute in (\ref{Rpi}) to obtain
{\footnotesize
\be
R(\tilde{q}^2)=-\frac{2\nf\nc}{(2\pi)^7 }{\mathrm{Im}}\left[ \frac{-2+2\sqrt{1+(\tilde{q}^2)}}{(\tilde{q}^2)}+H\left(\frac{1-\sqrt{1+(\tilde{q}^2)}}{2}\right)+H\left(\frac{-1+\sqrt{1+(\tilde{q}^2)}}{2}\right)\right],\label{Rren}
\ee }
where $H$ is the harmonic number function. We should notice here that $H(z)$ has an infinite set of singular points at $z=-k$ for $k\in{\mathbb{N}}^+$. These points are single poles with residue -1, making $R(\tilde{q})$ a series of resonances located at
\be
\tilde{q}_n=2\sqrt{n(n+1)}\,\,\,\,\,\,\, {\mathrm{for}} \,\,\,\,\,\,n=\left\{ 1,2,3,...\right\}. \label{poles}
\ee

We note that, although computed in a different way, the location of the resonances agrees with the spectrum of masses found in \cite{flavordavid} for the vector mesons classified therein as the $\ell=0$ case of the type II vector fields, which correspond to the ones we study here. The expression (\ref{poles}) is well defined for complex values of $\tilde{q}^2$, so it provides the analytic extension to the complex plane of the distribution obtained following the method in \cite{flavordavid}. 
 
Just to illustrate the effect of the quark mass in the spectrum prediction, we obtain the quark mass by fixing the first pole position to its experimental value through Eq. (\ref{poles}) ($n=1$, $N_c=3$), and predict the corresponding second pole position.  In Table \ref{spectrum} we compare the predicted and experimental second pole position for $u$, $s$ and $c$
 -quark  vector-meson states, using the $\rho(770)$, $\phi(1020)$ and $J/\psi(3096)$ as the first poles, respectively.
\begin{table}[t]
\caption{Comparison of the second pole prediction against experimental measurements}
\begin{center}
\begin{tabular}{|c|c|c|}
\hline
Quark mass (GeV) & Prediction (GeV)& Measurement (GeV) \\
\hline
$M_u=0.266$&1.334&1.465\\
$M_s=0.352$&1.765&1.680\\
$M_c=1.069$&3.686&3.771\\
\hline
\end{tabular}
\end{center}
\label{spectrum}
\end{table}%
The prediction for the second pole of the $u$ channel agrees within $9\%$ with the experimental measurement, while for higher quark masses we find an agreement of around $5\%$.
 
Given that with the method used here we obtain an analytic expression, $R(\tilde{q}^2=\frac{g_s\nc}{\pi}\frac{q^2}{{m_q}^2})$, valid for $q^2$ over the complex plane, we can give a small but finite imaginary component $\epsilon$ to $q^2$ and take its real part to be equal to $s$, that is, $q^2=s+i\epsilon$. By doing this, it is possible to adjust $\epsilon$ to account for the finite width of the mesons, and once the value of $\epsilon$ has been fixed, $R\left( \tilde{q}^2=\frac{g_s\nc}{\pi{m_q}^2}(s+i\epsilon)\right)$ becomes a function of $s$, $R(s)$. This procedure is similar in spirit to the usage of the Breit-Wigner formula to describe unstable states in quantum field theory, suggesting that the imaginary part of the propagator could be seen as a product of the mass and decay width ($\Gamma$) of the state, namely,  $\Gamma\equiv\epsilon/\sqrt{\tilde{q}_n^2}$. In the large $\nc$ limit that we are working at, the mesons are fully stable and this is reflected on the fact that when $\epsilon\rightarrow 0$ we obtain a series of delta functions, which is consistent with what has been conjectured before for this way of computing the spectral function \cite{Myers:2008cj,Myers:2007we,brightbranes}.

\begin{figure}
\centerline{\epsfig{file=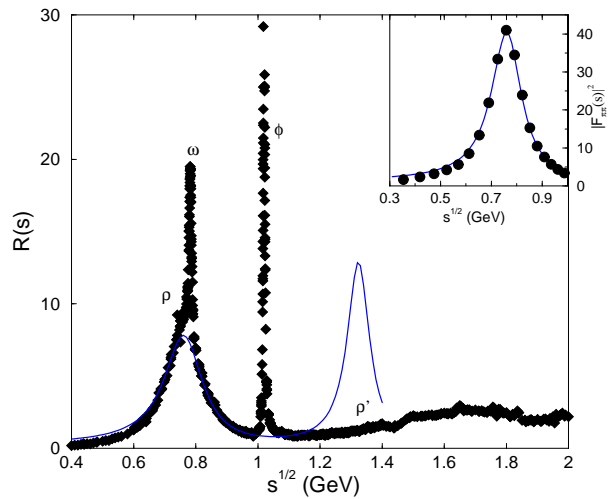,angle=-90,width=8cm}}
\vspace{-0.1in} \caption{ $R(s)$, prediction (solid line) and the measured values  from $e^+ e^-$ scattering (symbols). Upper panel, $R(s)$ predicted (solid line) and the experimental measurement from $\tau$ decay (symbols). See text for details.}\label{rplot}
\end{figure}

In order to have a qualitative insight on the predicted shape of $R(s)$, by including the decay width as discussed above, in Figure \ref{rplot} we plot $R(s)$ as obtained in the model (solid line) and the measured values (symbols), as obtained from $e^+ e^-$ scattering \cite{eedata}. 
We did choose the first pole to coincide with the $\rho(770)$ vector meson mass and decay width parameters, and the second pole position is predicted as we discussed previously. It is worth to mention that, since the width is not predicted, the width exhibited for the second pole is meaningless. It corresponds to having kept the same width as the one used for the $\rho(770)$ but substituting the predicted value for the mass of the second pole in the expression for the $\epsilon$ parameter. 
Notice that in our model, isospin is an exact symmetry and therefore degenerate isospin states  will not show as separate bumps. For the sake of clarity we have omitted to plot the pole associated to the $s$ quark state ($\phi$).

 Experimental information on $R(s)$ is also obtained via $\tau$ hadronic decays  by invoking isospin symmetry. In the upper panel of Figure \ref{rplot} we compare $R(s)$ (solid line) with the corresponding experimental measurements from $\tau$ data (symbols) \cite{taudata}. In each case we observe a good agreement between them. In particular, the shape is very well described. An obvious difference between the plots is the absence of the $\omega$ meson in the $\tau$ data, which appears in the $e^+ e^-$ data as a tall peak just above the $\rho$ peak. This is due to the fact that the former goes through a charged current channel.

To estimate the deviations in the shape of $R(s)$ compared to the experimental data, we compute its contribution to the anomalous magnetic moment of the muon $a_\mu^{had}$,
\begin{equation}
a_\mu^{had}(\pi\pi)=\frac{\alpha^2(0)}{3\pi^2}\int^\infty_{4m_\pi^2}ds\frac{K(s)}{s}R(s).
\end{equation}

This observable is suitable in this case, since the electromagnetic kernel $K(s)/s$ \cite{kernel} suppresses contributions for $\sqrt{s}\ge 1$ GeV and therefore only the first pole is relevant. We obtain $a_\mu^{had}(\pi\pi)=660 \times 10^{-10}$, which is about 25\% higher than the one obtained using experimental data (for example $520\times 10^{-10}$ from $\tau$ data \cite{taudata}). However, note that our value is not sharp, since the $R(s)$ normalization is not unique.

\section{Conclusions}
In summary, we have provided an analytic expression for $R(s)$ obtained using holographic techniques  at zero temperature in a strongly coupled, large-$\nc$ $SU(\nc)$ field theory with $\nf \ll \nc$ quark flavors. We saw that our result for $R(s)$ predicts the same mass spectrum for these kind of vector mesons as the one that had been established using other methods, and in addition it permits to include in a natural way the unstable character of the resonances. By comparison with the experimental data in the non-perturbative regime we can indeed recognize some of the gross features, as displayed in Figure \ref{rplot}; in particular there is good agreement on the shape of the first pole. Since we are working in a theory different from QCD, the localization of the second pole is not accurate enough but improves for higher quark masses. In addition, the prediction of the hadronic contribution to the anomalous magnetic moment of the muon is well approached. It is worth pointing out that our results cannot be used to check the high energy behavior since the action (\ref{sdq}) is dual only to the strongly coupled regime of the gauge theory.

\begin{acknowledgments} 
L.P. acknowledges support from PROFIP and PAPIIT IN-108309-3 (UNAM, M\'exico), and G.T. from CONACYT and DGAPA-PAPIIT project IN111609.
\end{acknowledgments}

\end{document}